\documentclass[12pt]{article} 

\usepackage {amsfonts,xypic} 

\author{J. Brian Pitts \\ Faculty of Philosophy \\ University of Cambridge \\ jbp25@cam.ac.uk}

\title{Space-time Constructivism \emph{vs.} Modal Provincialism: Or, How Special Relativistic Theories Needn't Show Minkowski Chronogeometry\footnote{To appear in the special issue of \emph{Studies in History and Philosophy of Modern Physics} on Harvey Brown's \emph{Physical Relativity} 10 years later.}}

\begin{document}   \sloppy
 \vspace{-.11in}

\maketitle

\begin{abstract}

Already in 1835 Lobachevski  entertained the possibility of multiple (one might say ``rival'') geometries of the same type playing a role.  This idea of rival geometries has reappeared from time to time (including Poincar\'{e} and several 20th century authors) but had yet to become a key idea in  space-time philosophy  prior to Brown's \emph{Physical Relativity}.  Such ideas are emphasized towards the end of Brown's book, which I  suggest as the interpretive key.  A crucial difference between Brown's constructivist approach to space-time theory and orthodox ``space-time realism'' pertains to modal scope.  Constructivism takes a broad modal scope in applying (at least) to all local classical field theories---modal cosmopolitanism, one might say, including theories with multiple geometries.  By contrast the  orthodox view is modally provincial in assuming that there exists a \emph{unique} geometry, as the familiar theories (Newtonian gravity, Special Relativity, Nordstr\"{o}m's gravity, and Einstein's General Relativity) have. These theories serve as the ``canon'' for the orthodox view.  Their  historical roles also suggest a  Whiggish story of inevitable progress.  Physics literature after \emph{c.} 1920 is relevant to orthodoxy primarily as commentary on the canon, which closed in the 1910s.  The orthodox view explains the spatio-temporal behavior of matter in terms of the manifestation of the real geometry of space-time, an explanation works fairly well within the canon.  The orthodox view, Whiggish history, and the canon have a symbiotic relationship.    

If one happens to philosophize about a theory outside the canon,  space-time realism sheds little light on the spatio-temporal behavior of matter.   Worse, it  gives the \emph{wrong} answer when applied to an  example arguably \emph{within} the canon, a sector of Special Relativity, namely, \emph{massive} scalar gravity with universal coupling.  Which is the true geometry---the flat metric  from the Poincar\'{e} symmetry group, the conformally flat metric exhibited by material rods and clocks, or both---or is the question faulty?  How does space-time realism explain the fact that all matter fields see the same curved geometry, when so many ways to mix and match exist?  Constructivist attention to dynamical details is vindicated;  geometrical shortcuts can disappoint.  The more exhaustive exploration of relativistic field theories in particle physics, especially massive theories, is a largely untapped resource for space-time philosophy.  

\end{abstract}



\section{Introduction}


Already in 1835 Lobachevski proposed that there might be more than one geometry that plays a role in one and the same real world. 
\begin{quote} We cognize directly in nature only  motion, without which all the impressions our senses receive become impossible.   All other ideas, for example geometric, though tied up implicitly in the properties of motion, are artificial products of our minds; and consequently space, by its own self, abstractly, for us does not exist. After that, there is no contradiction in our mind when we admit that certain natural forces obey the laws of a certain geometry, while others are governed by the laws of another geometry particular to them. \cite[Russian original 1835]{LobachevskiNewPrinciples} 
\end{quote} 
 It might be worthwhile to present another translation of a portion of this passage:  ``\ldots there can be no contradiction if we assume that certain forces in nature follow one geometry, and others a different geometry'' \cite{LogunovGeometrized}.  The idea of multiple geometries in one theory is a crucial idea that alters the big picture of space-time philosophy and is a matter of current mainstream physics research \cite{SolomonBigravityFinsler}. 

It turns out that Lobachevski's book was translated into English and published already in the 19th century \cite{LobachevskiNewPrinciples}, although not in the most prominent location (the Neomon, Austin, Texas, across the street from the university).  Translator George Bruce Halsted later translated Poincar\'{e}'s works \cite{PoincareFoundations}.  Halsted's comments in the ``Translator's Preface'' (appearing at the end of the work, judging by the Yale scan) show his recognition of the value of this portion of Lobachevski's idea here:
\begin{quote} 
it is preeminently in his ``New Principles'' that the great
Russian allows free expression to his profound philosophic insight,
which on the one hand shatters forever Kant's doctrine of
our absolute \emph{a priori} knowledge of fundamental spatial properties,
while on the other hand emphasizing the essential relativity
of space.
Lobachevski's position is still after sixty years the final
philosophy.
No one has gone beyond it.   \cite{LobachevskiNewPrinciples} \end{quote} 
Curiously enough, Lobachevski still has something to teach has not sixty but 180 years later.  
\begin{quote} The possibility of matter coupling to two metrics at once is considered.  \ldots matter can couple to two distinct metrics. Consequently, the traditional notion of a `physical metric' may have to be discarded, leaving us faced
with entirely new conceptual challenges in interpreting even the observables of the
theory.  \cite{SolomonBigravityFinsler}
\end{quote} 

A search of WorldCat suggests that this translation was not reproduced widely enough to make much impact. Evidently there is one copy outside the USA (University of London) besides the ten copies in the USA (Arizona, Southern California, Yale, Boston Athenaeum, Brandeis, New Mexico State,  Columbia, Syracuse, Brown, and, naturally,  the University of Texas at Austin).  It is easy to see how such a work could stay obscure, though I haven't ruled out the possibility that there is some other means of access to this idea of Lobachevski's.   Fortunately Yale has digitized the work and put it online.

Until recently  the possibility of rival geometries, multiple geometric structures of the same type (\emph{e.g.}, metrics, volume elements, or the like), seemed to originate first with Poincar\'{e} \cite[p. 489]{BenMenahemPoincare}  \cite{ScalarGravityPhil}.  The key passage, fairly familiar but easy to overlook, runs as follows:
\begin{quote}
Suppose, for example, that we have a great sphere of radius $R$ and that the temperature decreases from the center to the surface of this sphere according to the law of which I have spoken in describing the non-Euclidean world.

We might have bodies whose expansion would be negligible and which would act like ordinary rigid solids; and, on the other hand, bodies very dilatable and which would act like non-Euclidean solids.  We might have two double pyramids $OABCDEFGH$ and $O^{\prime}A^{\prime}B^{\prime}C^{\prime}D^{\prime}E^{\prime}F^{\prime}G^{\prime}H^{\prime}$ and two triangles $\alpha\beta\gamma$ and $\alpha^{\prime}\beta^{\prime}\gamma^{\prime}$. The first double pyramid might be rectilinear and the second curvilinear; the triangle $\alpha\beta\gamma$  might be made of inexpansible matter and the other of a very dilatable matter.

It would then be possible to make the first observations with the double pyramid $OAH$ and the triangle $\alpha\beta\gamma$, and the second with the double pyramid   $O^{\prime}A^{\prime}H^{\prime}$ and the triangle $\alpha^{\prime}\beta^{\prime}\gamma^{\prime}$.  And then experiment would seem to prove first that the Euclidean geometry is true and then that it is false.

\emph{Experiments therefore have a bearing, not on space, but on bodies.}  \cite[pp. 88, 89]{PoincareFoundations} (emphasis in the original) 
\end{quote}

Crucially, the idea in both Lobachevski and Poincar\'{e} is that both geometries actually do some observable work directly; more specifically, there is some type of  matter (Lobachevski) or token of matter (Poincar\'{e}) that exhibits each geometry chronogeometrically. Such works contrast with another use of a bimetric formalism, namely in Nathan Rosen's modification of General Relativity \cite{Rosen1}, which posits a flat background metric that is unobservable.\footnote{Curiously, the second of the two Rosen papers seems to envisage some local empirical consequences; that suggestion was mistaken.}  Such merits as Rosen's idea has are more subtle, especially  providing resources for engineering a consistent notion of causality, avoiding a chicken-and-the-egg problem that afflicts quantum gravity \cite{NullCones1}. Chris Isham has commented on this problem:
\begin{quote}
	  For example, in the Wheeler-DeWitt approach, the configuration variable of the system is the Riemannian metric $q_{ab}(x)$ on a three-manifold $\Sigma,$ and the canonical commutation relations invariably include the set 
\begin{equation}
[\hat{q}_{ab}(x), \hat{q}_{cd}(x^{\prime} )] = 0
\end{equation}
for all points $x$ and $x^{\prime} $ in $\Sigma.$  In normal canonical quantum field theory such a relation arises because $\Sigma$ is a space-like subset of spacetime, and hence the fields at $x$ and $x ^{\prime}$ should be simultaneously measurable.  But how can such a relation be justified in a theory that has no fixed causal structure?  The problem is rarely mentioned but it means that, in this respect, the canonical approach to quantum gravity is no better than the covariant one.  It is another aspect of the `problem of time' \ldots. \cite{IshamPrima} (p. 12)
\end{quote}
 But there is at least one other way to have two metrics in a theory, one that arises naturally in some potentially fundamental physics theories,  in which all matter sees one geometry, but the gravitational laws subtly display a different geometry, as will appear below.

Differential geometry was still quite immature when Einstein developed General Relativity.  For example, the connection had not yet been identified as in principle independent of a metric, potentially containing properties that go beyond a metric (torsion) or exhibiting a relation of not really fitting with some particular metric (non-metricity \cite{Schouten}) and perhaps not with any metric \cite{Edgar1,Edgar2,ScalarGravityPhil}.  As differential geometry matured in the late 1910s and 1920s, it was recognized by Levi-Civita that, \emph{e.g.}, there was no reason that a manifold must have only one metric.   
\begin{quote}  
There is clearly no reason against assigning in turn to the same analytical manifold two distinct metrical determinations,
defined by the two quadratic forms [footnote suppressed]
$$ds^2 = {\sum_1^n}_{ik}  a_{ik}dx_i dx_k, . . . (1) $$
$$ds^{\prime2} = {\sum_1^n}_{ik} a^{\prime}_{ik} dx_i dx_k, . . . (1')$$  \cite[p. 220]{LeviCivita}.
\end{quote}
If there is no reason that a manifold should have only one metric, then there is no reason not to entertain such possibilities in physical theorizing.  


\section{Two Metrics, One Directly Observable, One Indirectly} 

An interesting possibility, plausible in fundamental physics, is when there are two geometries in a theory, one of them encodes the behavior of (all) rods and clocks, and the other is \emph{indirectly} observable, though not because any rods or clocks give it chronogeometric significance.   One generally doesn't expect to find deep philosophy of geometry in the \emph{Astrophysical Journal}, but one should be glad for exceptions.  
A \emph{prima facie} plausible philosophy of geometry for bimetric massive variants of (\emph{i.e.}, rivals to) General Relativity was outlined clearly by Freund, Maheshwari and Schonberg in the late 1960s in connection with their massive spin-$2$ gravitational theory \cite{FMS}.  It differs from Einstein's theory by adding a graviton mass term, a part in the Lagrangian density involving algebraic quadratic (and perhaps higher) terms in the gravitational potential. The field equations thus receive a linear algebraic term (and perhaps higher terms).  The gravitational potential is basically the difference between the metric $g_{\mu\nu}$ and some background metric $\eta_{\mu\nu}.$\footnote{In fact Freund, Maheshwari and Schonberg in fact used the matrix $diag(-1,1,1,1)$ rather than a flat background metric tensor.  Thus they avoided  formal general covariance in favor of preferred  Cartesian coordinates adapted to the flat background.  Introducing a flat background metric \emph{tensor} $\eta_{\mu\nu}$ (with a component transformation law),  not just a matrix $diag(-1,1,1,1)$, puts the theory in a more familiar form for foundational purposes and exposes two metrics in the theory.} 
 The graviton mass term takes the form 
\begin{equation}
\frac{m^2}{16\pi G}\int d^4x  (\sqrt{-g} +\sqrt{-\eta} - \frac{1}{2} \sqrt{-g} g^{\mu\nu}\eta_{\mu\nu} ),
\end{equation}
One could make a perturbative expansion to express this mass term in terms of quadratic and higher terms in the gravitational potential.   I will not do so here, partly because such expansions introduce questions of field redefinitions that tend to obscure the empirical/chronogeometric significance of the quantities.\footnote{Should one define the gravitational potential using  $g_{\mu\nu}-\eta_{\mu\nu}$, or $\frak{g}^{\mu\nu} - \sqrt{-\eta} \eta^{\mu\nu}$ where $\frak{g}^{\mu\nu} = \sqrt{-g}g^{\mu\nu},$ or what  \cite{DeserMass}?  Of course fundamentally it doesn't matter as long as one keeps track of what one means.  There are practical advantages for one choice or another in calculations, but these vary with context:  $g_{\mu\nu}-\eta_{\mu\nu}$ leads to simpler  Hamiltonian treatments, whereas $\frak{g}^{\mu\nu} - \sqrt{-\eta} \eta^{\mu\nu}$ leads to simpler wave equations \cite{Papapetrou}, and certain fractional-weight definitions make the vacuum General Relativity Lagrangian polynomial \cite{DeWitt67b}, namely, $(-g)^{\frac{5}{18}}g^{\mu\nu} - (-\eta)^{\frac{5}{18}}   \eta^{\mu\nu} $ or   $(-g)^{-\frac{5}{22}}g_{\mu\nu} - (-\eta)^{-\frac{5}{22}}   \eta_{\mu\nu}$.  How one would express DeWitt's quantities using the resources of modern geometry is worth reflection.  Most other choices have no particular advantages in any context.}

Freund, Maheshwari and Schonberg were very clear about many aspects of the physical meaning of their two metrics:
 \begin{quote} 

\hspace{1.25in}  a) \emph{Breakdown of Geometrical Interpretation} \\ 

The theory, not being generally covariant, cannot be interpreted geometrically. This
means first of all that the quadratic form,
$$ d\sigma^2 = g_{\mu\nu} dx^{\mu} dx^{\nu} \,\,, $$
has nothing to do with the line element of the \emph{world} geometry, which remains
$$ds^2 = \eta_{\mu\nu}dx^{\mu} dx^{\nu} \,\, .$$

Similarly, the equations of motion of matter (25) still look formally as if they were geodesic
equations. As a matter of fact, they are not. Indeed, the $\Gamma_{\mu\nu}^{\sigma}$ are given by the usual
expressions, but $g_{\mu\nu}$ and\ldots  [its  inverse] are determined from the \emph{not}-generally-covariant equations (28), so that the $\Gamma_{\mu\nu}^{\sigma}$ are not genuine Christoffel symbols. The geometrical interpretation is one of the crucial steps in applications of Einstein's theory. What do we offer as a replacement? The field equations (28) and the equations of motion for matter (25) fully
determine the answer to any question one can ask. For that matter, this is true for
Einstein's theory as well. There, however, geometrical considerations may be used as a
luxurious shortcut toward the answers to many problems.

 \hspace{2in}   b) \emph{Local Problems} \\ 
If our theory is different from Einstein's, does this mean that it conflicts with the classical
tests of the latter? No. All classical tests are \emph{local}, i.e., they involve only small
regions of space and time. Locally our theory differs from that of Einstein only by
terms of the order (radius of system/Hubble radius), so that the corrections are indeed
negligible and the local tests cannot distinguish between the two theories. Moreover,
\emph{locally} one can reinstate an approximate geometrical interpretation.\ldots 
Thus at a local level our theory is indistinguishable by usual experiments from that of Einstein. The real
difference appears for systems of the size of $\Lambda^{-\frac{1}{2}},$ that is, for cosmological problems.
 \cite{FMS}
\end{quote}

Before long it was argued that important problems of detail afflict not only the Freund-Maheshwari-Schonberg theory, but all similar theories  \cite{vDVmass1,vDVmass2,DeserMass}. Some of those arguments were later criticized   \cite{Vainshtein2,deRhamGabadadze,HassanRosen}, with ongoing debate \cite{DeserWaldronAcausality}.
Such issues are relevant to whether it is epistemically possible for us now that such a theory is true of the actual world.  If one is interested only in whether theories with similar features are metaphysically possible, and epistemically possible as of the 1910s rather than 2016, then one can use a massive \emph{scalar} theory instead \cite{UnderdeterminationPhoton,PittsScalar,ScalarGravityPhil}, doing to Nordstr\"{o}m's theory what Hugo von Seeliger and Carl Neumann in the 1890s had done and Einstein in 1917 \cite{EinsteinCosmological} would do to Newton's theory.  Hence the general philosophical idea of Freund, Maheshwari and Schonberg manifestly could be realized for scalar gravity theories, whether or not it could be for tensor theories such as they considered.

Related  points have been made in the philosophical literature   \cite{WeinsteinScalar} using Brans-Dicke scalar-tensor gravitational theory. In such theories, there is one metric that rods and clocks exhibit chronogeometrically (the ``Jordan frame,'' because the theory is then formulated much as in  Jordan's original work before Brans-Dicke), whereas a different metric (the ``Einstein frame,'' conformally related to the other one) provides a simple picture of the gravitational dynamics.  The physics literature at times has suffered from a debate about which metric is the real metric. Some authors prudently made the conventionalist proposal that there is no fact of the matter \cite{FaraoniNadeau}.   Why would such a question need a context-independent answer?


\section{Reading \emph{Physical Relativity} Starting at the End} 

Presumably many readers  read Brown's \emph{Physical Relativity} starting at the beginning.  That is the natural thing to do with books, especially for diligent readers who plan to read the whole book.  I suspect that most reviewers have read the book the same way. 

 While this style of reading is natural, it has a disadvantage in approaching a book such some of the key points for good reasons come at the end.  \emph{Physical Relativity} is such a book.  In terms of both chronology and difficulty, one comes to Special Relativity before General Relativity.  While one might recognize the importance of a dynamical approach based on studying General Relativity or even other theories, one might still feel constrained to write about Special Relativity first.  It isn't so easy to make clear why one \emph{must} take a dynamical/constructive approach to Special Relativity, though Brown does  a good job arguing the point.  But once one has seen the big picture from the end of the work, where the discussions of non-minimal coupling in General Relativity (pp. 165-172) and of multi-geometry theories such as Bekenstein's TeVeS (pp. 172-176), it becomes clear that the point is generic.  Hence Special Relativity is a \emph{degenerate case}.    
In fact all the familiar  cases on which space-time philosophy dwells, Newton, Special Relativity, (if the writer has historical interests) Nordstr\"{o}m's scalar theory, and General Relativity, are all examples of a happy degeneration, namely, there are no rival geometrical structures  in their usual formulations.\footnote{One noteworthy exception \cite{EarmanWorld}  is organized not so much around actual theories as around aspirations for theories during the 17th-19th centuries. }    Space-time has a unique spatial metric and a unique temporal metric (or a unique spatio-temporal metric), and a unique affine connection (though admittedly there is an issue of choice regarding Newtonian gravity, 
still one doesn't need both  at the same time).

My own way of reading Brown's work was initially more a matter of reading the table of contents for parts that sounded especially fascinating (such as the last chapter, 9, ``The View from General Relativity,'' and the adjacent appendix A ``Einstein on General Covariance'' and some other parts).  This approach, while not ideal in general, in this case had the advantage of making it easier to view the whole with an emphasis on what I take to be key points made near  the end.   By contrast generally reviews have not had much to say about the last chapter and first appendix. %
While failing to criticize perhaps the best part of the book counts for something, one could wish for recognizing the best part rather than passing over it in silence.  Likely reviewers' opinions of the book were largely fixed by the time they reached these last parts of the book.  The entrenchment of the canon in space-time philosophy is so strong that the concluding parts of the book were apparently too little, too late for most reviewers to overcome  modal provincialism, the assumption that the whole realm of interesting theories that space-time philosophy should consider is basically like the familiar theories.  In my view, issues of non-minimal coupling and theories with rival geometries should govern one's overall view in order to achieve a modally cosmopolitan view. %


\section{Should Space-time Philosophy Have a Canon? }

At least until recently, English literature  was thought to have a canon, a collection of the most important works that everyone should know and that provide a standard against which all others can be measured.  Evidently this fact itself is not as old as it might seem.  But the idea of a literary canon and its analogs provide a useful model for contemplating space-time philosophy.  
\begin{quote} 
The normative sense of canon has been strongly reinforced by the
application of the term to the accepted books of the Bible, though there
is no agreement on the original force of the word even in this application. [footnote suppressed]  %
  \cite{Canonicity}  \end{quote}  
Harris writes in a field where the idea of a literary canon is very familiar, albeit contested in various ways, and where the Biblical parallel has been arguably overstated.  
\begin{quote}  
Though the sense of ``unquestionably and
uniquely authoritative'' that belongs to the biblical
canon (and to the theologically derived endorsements
and prohibitions churches enforce on members of their faith) continually colors the debate
over the modern literary canon, the analogy
is more dramatic than helpful. \cite{Canonicity} \end{quote}  
But are there fields in which a canon operates \emph{de facto} but unrecognized and where calling attention to the parallel is illuminating?

The utility of calling attention to a \emph{de facto} canon is all the more plausible if a field contains a prophet(s).  
Pais's scientific biography of Einstein is useful in that respect.  Pais  finds 
``the parallels with the rituals of beatification and canonization compelling\ldots'' and portrays a joint meeting of the Royal Society and the Royal Astronomical Society on November 6, 1919 as a canonization ceremony \cite[p. 305]{PaisEinstein}.
This behavior of two royal societies was, of course, not just the extravagance of the popular press, though the latter was important.  Pais attempts both to describe and explain the place that Einstein came to occupy.  
\begin{quote}
The essence of Einstein's unique position goes deeper and has everything to do,
it seems to me, with the stars and with language. A new man appears abruptly,
the `suddenly famous Doctor Einstein.' He carries the message of a new order in
the universe. He is a new Moses come down from the mountain to bring the law
and a new Joshua controlling the motion of heavenly bodies. He speaks in strange
tongues but wise men aver that the stars testify to his veracity.\ldots 
Behold, a new man appears. His mathematical language is sacred yet amenable
to transcription into the profane: the fourth dimension, stars are not where they
seemed to be but nobody need worry, light has weight, space is warped. He fulfills
two profound needs in man, the need to know and the need not to know but to
believe. The drama of his emergence is enhanced (though this to me seems secondary) by the coincidence---itself caused largely by the vagaries of war---between
the meeting of the joint societies and the first annual remembrance of horrid events
of the recent past which had caused millions to die, empires to fall, the future to
be certain. The new man who appears at that time represents order and power.
He becomes the \ldots  the divine man, of the twentieth century.
 \cite[p. 311]{PaisEinstein}    

\end{quote}

Is all this hagiography something 
 above which philosophers have risen?   Consider any famous physicist besides Einstein (or Newton in his day)---say, Pauli, Landau, Feynman, Dirac, or the like.  One will not find nearly the same kind of anti-anti-Pauli, or anti-anti-Landau, \emph{etc.} reflex if the idea of disagreeing or moving beyond one of these giants is entertained. While such criticisms have a good chance of being unwise, they are not nearly so automatically embarrassing and shameful.   It is doubtless relevant that  Einstein endured unusually much criticism that was stupid and/or malicious, some if it from philosophers, partly because he was famous and his work touched on issues of widespread interest with entrenched views.  While we do not want to be associated with stupid or malicious criticism (embarrassing ourselves and the  discipline), it is not necessary in 2016 to continue to prove our credentials (that unlike Bergson and Dingler, we are free of such defects as unrevisable \emph{a priori}  philosophy of time and/or space and, in the latter case, ties to National Socialism) by attending to Einstein's theory almost exclusively as a lasting and perpetual achievement \emph{far beyond what physicists have been doing for the last 15+ years in the wake of dark energy}.  Besides huge amounts of work in the physics literature that could lead to replacing General Relativity,  in some cases with a merely Poincar\'{e}-invariant theory (\emph{e.g.}, \cite{deRhamGabadadze,HassanRosen}) rather than a generally covariant one, there has been since 2009 even substantial discussion of the idea that one can achieve a \emph{renormalizable theory of quantum gravity by rejecting even Special Relativity in favor of absolute simultaneity} \cite{Horava}.  This work has been cited well over 1000 times. If one does not like these examples, even scalar-tensor theories, presumably as covariant as anyone could want, make the same multi-geometry point \cite{WeinsteinScalar}, which nevertheless keeps getting lost.   Will philosophers acquire the self-confidence to stop trying to prove our orthodoxy more zealously than the scientists who know at least as much about space-time as we do?  Brown 
reached that stage in 2005 by philosophizing about extra-canonical works such as TeVeS.


Some related developments are worth mentioning. A number of the leading historians of General Relativity, after looking at Einstein's unpublished as well as published works through 1915, have concluded that Einstein found the definitive field equations by  a more physical and less mysterious process than has been thought    \cite{RennSauer,Janssen,RennDwarfEmergence,Renn,JanssenRenn,RennSauerPathways}. This long-neglected ``physical strategy'' bears a strong resemblance to the later particle physics tradition \cite{EinsteinEnergyStability}.   
  Einstein apparently re-wrote his own history partly in order to justify his decreasingly appreciated unified field theory quest \cite{vanDongenBook}. 
More broadly focussed work now places Einstein within his generation \cite{StaleyGeneration}.  The idea that physics is being distorted by learning the wrong lessons from faulty history is being suggested by a leading physicist \cite{SmolinEinsteinDiscovery}.  Finally, Einstein's treatment of conservation laws and their connection to symmetries, far from being prescient, was not even viable or well informed of old or recent results \cite{EinsteinEnergyStability}.  Hence  a variety of authors have recently worked toward de-privileging the mystical (and mythical) Einstein.  The propriety of a  dynamical approach to space-time philosophy is thus confirmed from other directions.


\section{Norton's Space-time Realism}  

One review of \emph{Physical Relativity} deserves special attention \cite{NortonFails}.  That is partly because its title ``Why Constructive Relativity Fails'' might give the reader the impression that constructive relativity fails, and partly because it provides such a clearly stated alternative that can be evaluated.  Norton is quite correct that Brown does not construct every notion in sight, but how crucial is that  \cite{StevensDynamical}?
\begin{quote}
After all, the project was to reduce chronogeometric facts to symmetries, not to recover the entire spatiotemporal nature of the world from no spatiotemporal assumptions whatsoever. The constructivist's project might need a primitive notion of ``being contiguous'', but Norton is wrong to think that it follows from this that constructivists are illicitly committed to the independent existence of spacetime. [footnote suppressed] 
\cite[p. 573]{PooleyEncyclopedia}  \end{quote}
In any case Norton has provided a useful service  toward identifying a large core of overlap between constructivists like Brown and space-time realists.

Let us recall at some length what Norton calls the ``Realist conception of Minkowski spacetime''  \cite{NortonFails}.  (Evidently Norton takes this example to be sufficiently general to indicate how space-time realism would apply to other theories as well---a crucial assumption. The modest role for  physics literature after \emph{c.} 1920 for understanding  space-time realism is evident.)   
\begin{quote}
(1)  There exists a four-dimensional spacetime that can be coordinatized by a set of standard coordinates (x, y, z, t), related  by the Lorentz transformation. \\
(2)  The spatiotemporal interval $s$ between events $(x, y, z, t)$ and $(X, Y, Z, T)$  along  a straight [footnote suppressed]  line connecting  them  is a property  of the spacetime, independent  of the matter it contains, and is given by %
 $$ s^2 = (t - T)^2  - (x - X)^2  - (y - Y)^2 - (z - Z)^2. \hspace{.5in}	(1) $$
 When $s^2  > 0,$ the interval $s$ corresponds  to times elapsed on an ideal clock; when $s^2  < 0$, the interval $ s$ corresponds  to spatial  distances measured by ideal rods (both employed in the standard way). \\
(3)  Material  clocks and rods measure these times and distances because the laws of the matter  theories that  govern them are adapted  to the independent  geometry of this spacetime.

If constructivism  is to  be novel, it must  contradict this view; it certainly cannot covertly presume all or most of it.%

What I will seek to establish below is that if constructivism  is to succeed, it must tacitly presume at least a major part of this realist conception.  
\cite{NortonFails} \end{quote}
  
My  concerns are with (2) and (3).  Concerning (2), the definite article and singular noun in ``[t]he \ldots interval'' are worrisome. They suggest that Norton's analysis has not been honed with examples lacking One True Geometry in mind.  The space-time realist has an unfaced task, to give a plausible story outside the canon, as  Brown's constructivism clearly does \cite{BrownPhysicalRelativity,ButterfieldCausalityConventionGeometry}. %


\section{Massive Scalar Gravity \emph{vs.} Space-time Realism} 

More definitively troublesome is the fact, that Norton's (3) is \emph{false} even within what is presumably a sector of Special Relativity, namely, universally coupled massive scalar gravity, which is just Poincar\'{e}-invariant. These theories, though they have roots in the 1890s in the work of Seeliger and Neumann, and would have gravity satisfy the rather familiar Klein-Gordon equation in the lowest approximation, have hardly ever been studied thoroughly, especially for philosophical lessons, until recently  \cite{PittsScalar,UnderdeterminationPhoton,ScalarGravityPhil}. One cannot assume that physicists have already done all the work that philosophy requires.  
 The Freund-Nambu paper \cite{FreundNambu} was a milestone, though it doesn't actually discuss gravity and was connected with gravity only later by others \cite{DeserHalpern}. 
According to the universal coupling assumption, material stress-energy and gravitational stress-energy serve as sources for gravity in the same way.  That  idea was already advocated as part of Einstein's ``physical strategy''  in 1913.
\begin{quote} 
These equations satisfy a requirement that, in our opinion, must be imposed on a relativity  theory of gravitation; that is to say, they show that the tensor $\theta_{\mu\nu}$  of the gravitational field acts as a field generator in the same way as the tensor $\Theta_{\mu\nu}$ of the material processes.  An exceptional position of gravitational energy in comparison with all other kinds of energies would lead to untenable consequences. \cite{EinsteinEntwurf}  \end{quote} 

In the scalar case, only the \emph{trace} of stress-energy serves as such a source.  It is not obvious by inspection, but is nonetheless demonstrable, that as a result the matter field equations have matter coupled to an effective geometry that is curved (though conformally flat) because the gravitational potential merges with the original volume element \cite{Kraichnan,FreundNambu,DeserHalpern,PittsScalar,UnderdeterminationPhoton,ScalarGravityPhil}. 
Thus the field equations follow from a Lagrangian density of this form (with $u$ being matter, $g$ giving effective volumes \emph{via} the effective positive volume element $\sqrt{-g},$ $\hat{\eta}_{\mu\nu}$  being the unimodular Weyl-flat tensor density giving the conformal geometry, and $\eta$ being the determinant of the flat metric):
$$\mathcal{L}_{Nord}[ {g}, \hat{\eta}_{\mu\nu}] + \mathcal{L}_{mass}[{g}, {\eta}] +\mathcal{L}_{matter}[{g}, \hat{\eta}_{\mu\nu},u].$$
 The massive scalar gravitational potential is a volume-distorting \emph{almost}-universal force due to universal coupling.
 For massive theories,  $ {\eta}$ \emph{is observable in  long-range gravitational experiments}.
 So the chronogeometrically observable conformally flat metric $g_{\mu\nu}= \hat{\eta}_{\mu\nu}  (-{g})^{\frac{1}{4}}  $ isn't clearly the One True Geometry.
 The mass term both reduces the symmetry group from the 15-parameter conformal group of Nordstr\"{o}m's theory \cite{NortonNordstrom} to the Poincar\'{e} group and makes the undistorted volume element $\sqrt{-\eta}$ indirectly observable, hence not surplus.  Therefore one cannot eliminate it as one would eliminate  the original volume element in Nordstr\"{o}m's theory  or any analogous unobservable geometry that might have been posited to be hidden by Lotze-Poincar\'{e}-Reichenbach universal forces.

Presumably the controversial part of Norton's (3), the part disagreeing with Brown,  was intended to be the explanatory claim ``because the laws of the matter  theories that  govern them are adapted  to the independent  geometry of this spacetime.''  But massive scalar gravity provides an example where the empirical explanandum ``[m]aterial  clocks and rods measure these times and distances'' fails. 
 Whether an explanation requires a true explanans has been debated; one might think that Newtonian gravity is false and yet explains a lot about falling bodies, the solar system, \emph{etc.}  But an explanation with a false explanandum is another matter.  One doesn't need Minkowski space-time   to explain why rods and clocks exhibit the interval 
$$  s^2 = (t - T)^2  - (x - X)^2  - (y - Y)^2 - (z - Z)^2$$ 
when they actually don't. Instead they exhibit a position-dependent conformally flat geometry due to the gravitational potential.    (3) predicts that rods and clocks do something that contradicts what the field equations say that rods and clocks do, as one can see by inspection of the Lagrangian density $$\mathcal{L}_{Nord}[ {g}, \hat{\eta}_{\mu\nu}] + \mathcal{L}_{mass}[{g}, {\eta}] +\mathcal{L}_{matter}[{g}, \hat{\eta}_{\mu\nu},u],$$ because $\mathcal{L}_{matter}$ depends on $g$ and does not depend on $\eta$.  In such a theory, matter fields see an effective curved (conformally flat metric)  $g_{\mu\nu},$ not the flat background metric $\eta_{\mu\nu}$.  But $\eta_{\mu\nu}$ indeed is fully present and at least indirectly observable in the theory (and therefore cannot be removed as gratuitous like Lotze-Poincar\'{e}-Reichenbach universal forces)---one only needs to do sufficiently long-range gravitational experiments.  Relative to Norton's space-time realism, constructivism is indeed novel:  constructivism \emph{lacks the false explanandum} contained in (3) for the case of massive scalar gravity. In the case of massive scalar gravity, constructivism bests realism \emph{even within the canon}.

Brown's treatment of Special Relativity/Minkowski geometry/Poincar\'{e}-invariant theories (though perhaps one is now wary of assuming that these are all the same) requires only slight modification consistent with the main thrusts of his work.  He writes that 
\begin{quote} 
in special relativity, the Minkowskian metric is no more than a codification of the behaviour of rods
and clocks, or equivalently, it is no more than the Kleinian geometry associated
with the symmetry group of the quantum physics of the non-gravitational
interactions in the theory of matter. \cite[p. 9]{BrownPhysicalRelativity}  %
\end{quote} 
Massive scalar gravity lacks Minkowskian behavior of rods and clocks, though it has the Minkowski metric (among other things) and the Poincar\'{e} symmetry group.  Whereas massive scalar gravity undermines the core of Norton's view, it undermines an isolated sentence or two of Brown's work while showing that key points, that detailed dynamics is crucial for explanation and that it can lead to the failure of familiar geometric shortcuts, are even more broadly applicable than they seemed.

The question arises  what to do about theories that do not have One True Geometry.  This question certainly is discussed in Brown's book \cite{BrownPhysicalRelativity}.   Strikingly, the question often does not arise  in responses to the book,   
 much as it failed to arise in Eddington's response to Poincar\'{e} 90-odd years ago \cite{EddingtonSTG,PoincareFoundations}.  Given the emphasis placed on such possibilities in Brown's book, responding by  reverting to Special Relativity prevents one from interacting with one of the book's strongest points.  Perhaps reviewers have no objection to the  treatment of such theories (though how such a view would fit with the treatment of the canon is difficult to say), or perhaps the argument has not been highlighted sufficiently to make its importance evident, as this paper aims to do (as has another work by Brown \cite{BrownRods}).

  In general, geometry isn't very illuminating, restrictive, or explanatory; when it appears to be, an assumption of few ingredients has been introduced.  Geometry is no royal road to physics with sufficient modal scope for the philosophy of geometry.   Space-time realism works, when it does, by assuming a special case, starting near the finish line.
 In that respect space-time realism reminds one of Bertrand Russell's remark:  
\begin{quote} 
The method of ``postulating'' what we want has many advantages; they are the same as the advantages
of theft over honest toil. Let us leave them to others and proceed with our honest toil. \cite[p. 71]{RussellMathematical} \end{quote} 
   Space-time realism apparently has nothing to say about extra-canonical theories lacking One True
Geometry. It is modally provincial. By contrast constructivism motivated partly by taking seriously the possible non-existence of One True Geometry. It is modally cosmopolitan, much like Lobachevski and like Poincar\'{e}'s and some subsequent authors' conventionalism \cite{LobachevskiNewPrinciples} \cite[pp. 88, 89]{PoincareFoundations} \cite{BenMenahemPoincare,GrunbaumEarman,WeinsteinScalar,FMS}.  
  Constructivism heeds Norton's good advice not to impart spurious necessity to contingent claims about best theory, General Relativity  \cite[pp. 848, 849]{Norton}, better than  space-time realism does.

There seems to be no natural and attractive way to amend space-time realism to fix this problem.  If one decides (likely contrary to the taxonomic inclination of most physicists) that Poincar\'{e}-invariant massive scalar gravity isn't really part of Special Relativity, then it turns out that space-time realism  isn't very good at figuring out which theories were part of Special Relativity, and needs particle physics/constructivist/conventionalist examples to make that point, which is awkward. The space-time realist's need to address theories lacking One True Geometry is still unresolved even if the taxonomic exclusion is made, because massive scalar gravity, even if not part of Special Relativity or a worthy participant in Minkowski space-time, still could have been true.  Indeed it was epistemically possible and not terribly unlikely for a time in the mid-1910s  that it was actually true.  Any adequate philosophy of geometry will have to borrow a page (perhaps a chapter) from the constructivist's attention to detail and a broad range of possibilities and rip out a  page about the adequacy of geometric shortcuts. Poincar\'{e} \cite{PoincareFoundations} could have said the  same thing in reply to Eddington \cite{EddingtonSTG}.  But Poincar\'{e} was dead, and no one said it on his behalf.
One needs not just a list of geometric objects, but what they \emph{do} in  field equations. With a single metric, General Relativity (or Nordstr\"{o}m's theory if conformally flat) follows uniquely (if higher derivatives are excluded); but with more ingredients, there are more possibilities.  As with cooking, so with differential geometry:  with more than one ingredient, it becomes necessary to say something detailed  about proportions.

One of the main merit's of Brown's work is that his views both arise from and reinforce the habit of paying attention to extra-canonical literature.  Thus he has been able to articulate as a core theme a point that only emerged sporadically earlier, namely, the need to pay attention  to a broad range of examples, thereby to recognize how the handful of usual cases attended in philosophy are unrepresentative special cases, to foreground the need to pay attention to details, and, in part, to show how that philosophy applies {even  for a proper understanding of at least some of  those special cases}.


\section{Priority of Field Equation(s) over Geometry}

One of the key points at issue between constructivism and orthodox space-time realism is the question of the relative explanatory priority of geometry (the orthodox view) or the dynamical laws (the dynamical/constructivist view).  In General Relativity, Riemannian  geometry greatly restricts  field equations.   Perhaps geometry \emph{explains} the field equations?   General Relativity is the {only} option given just \emph{a} metric  (assuming general covariance, without which one could hide other entities using preferred coordinates) and without higher derivatives.   Likewise for Nordstr\"{o}m-Einstein-Fokker (massless) scalar gravity, the conformally flat geometry gives only one choice, hence arguably explaining the field equations.     What is the One True Geometry for massive scalar gravity? Is it the flat metric $\eta_{\mu\nu}$ giving the Poincar\'{e} symmetries of Special Relativity?  90\% of it is chronogeometrically observable, while the remaining 10\% (volumes) are at least indirectly observable using long-range gravitational observations.  Or is the One True Geometry the curved (conformally flat)  metric $g_{\mu\nu}$ manifested materially by  rods and clocks, which are made of matter fields $u$?
Is it both?  No answer explains the field equations, adequately summarized by the schematic Lagrangian
$$\mathcal{L}_{Nord}[g_{\mu\nu}] + \mathcal{L}_{mass}[{g}, {\eta}] +\mathcal{L}_{matter}[g_{\mu\nu},u].$$
The mass term{\bf \underline{s}} (there is not just one as in \cite{FreundNambu}, but many, much like the spin $2$ case \cite{OP} at least \emph{prima facie}) are
$$  \mathcal{L}_{mass} =  \frac{m^2}{64 \pi G} \left[ \frac{  \sqrt{-g} }{w-1}   +   \frac{ \sqrt{-g}^{\ w} \sqrt{-\eta}^{1-w} }{w(1-w)}  -  \frac{ \sqrt{-\eta}  }{w} \right] $$  \cite{PittsScalar,UnderdeterminationPhoton}. 
Even if one accepts that as the One True Geometry a bimetric geometry, that doesn't explain the \emph{specific} $w$, or why matter $u$ sees no hint of the original volumes $\sqrt{-\eta}.$ The latter is explained by universal coupling.  It is unclear that anything but the field equation itself  can explain why one specific $w$ is selected.  But the explanatory priority of the field equations over geometry is just what Brown was at pains to show.  Hence massive scalar gravity is a good example for the constructivist claim that the field equations are explanatorily prior to the geometry.


\section{Conclusion}

If one takes the engagement with non-minimal coupling and multi-geometry theories at the end of Brown's book as the interpretive key, then one finds a fundamental issue that has only intermittently been recognized previously in space-time philosophy.  The issue is crucial because otherwise space-time philosophy is simply mute when confronted with many theories that could have been true, or even might yet be true.  Critiques of Brown's work have tended to ignore multi-geometry theories.  The new example of massive scalar gravity shows that space-time realism doesn't work well   \emph{even within the canon}.  The possible multiplicity of geometries in a theory and resulting greater multiplicity of how the ingredients can be employed shows that laws are indeed explanatorily prior to geometry.  The key example of massive scalar gravity is the sort of theory that doesn't arise at all readily within General Relativity-shaped imaginations but arises very  easily within imaginations expanded by particle physics.  Hence particle physics is a key neglected ingredient for space-time philosophy.  


\section{Acknowledgments}

I thank Harvey Brown for discussions of his work and  its reception and Simon Saunders for the invitation to submit to this special issue. This work was funded by the John Templeton Foundation,  grant \# 38761.



\end{document}